\documentclass{article}

\PassOptionsToPackage{numbers, compress}{natbib}
 \usepackage[preprint]{neurips_2025}

\usepackage[utf8]{inputenc} 
\usepackage[T1]{fontenc}    
\usepackage{hyperref}       
\usepackage{url}            
\usepackage{booktabs}       
\usepackage{amsfonts}       
\usepackage{nicefrac}       
\usepackage{microtype}      
\usepackage{xcolor}         
\usepackage{multirow}
\usepackage{makecell}
\usepackage{tabularx}
\usepackage{ltablex}
\usepackage{adjustbox}
\usepackage{array}
\usepackage{graphicx}
\usepackage{caption}
\usepackage{float}
\usepackage{gensymb}
\usepackage{amsmath}

\newcolumntype{Y}{>{\centering\arraybackslash}X}

\keepXColumns
\title{Discontinuous Epitope Fragments as Sufficient Target Templates for Efficient Binder Design}
\workshoptitle{AI for Science}
\author{%
Zhenfeng Deng$^{1,2}$  \quad  Ruijie Hou$^{2}$ \quad  Ningrui Xie$^{1,2}$ \quad \\ \textbf{Mike Tyers}$^{1,2}$\thanks{Correspondence to \texttt{michal.koziarski@sickkids.ca} and \texttt{mike.tyers@sickkids.ca}} \quad \textbf{Michał Koziarski}$^{1,2,3,4}$\footnotemark[1] \\ %
$^1$University of Toronto, \quad $^2$The Hospital for Sick Children Research Institute, \\
$^3$Vector Institute, \quad $^4$ Acceleration Consortium\\\\
\texttt{\{zfevan.deng,ningrui.xie\}@mail.utoronto.ca}\\
\texttt{\{ruijie.hou,michal.koziarski,mike.tyers\}@sickkids.ca} %
}

\begin{document}

\maketitle

\begin{abstract}
Recent advances in structure-based protein design have accelerated \textit{de novo} binder generation, yet interfaces on large domains or spanning multiple domains remain challenging due to high computational cost and declining success with increasing target size. 
We hypothesized that protein folding neural networks (PFNNs) operate in a “local-first” manner, prioritizing local interactions while displaying limited sensitivity to global foldability. 
Guided by this hypothesis, we propose an epitope-only strategy that retains only the discontinuous surface residues surrounding the binding site. 
Compared to intact-domain workflows, this approach improves \textit{in silico} success rates by up to 80\% and reduces the average time per successful design by up to forty-fold, enabling binder design against previously intractable targets such as ClpP and ALS3. 
Building on this foundation, we further developed a tailored pipeline that incorporates a Monte Carlo–based evolution step to overcome local minima and a position-specific biased inverse folding step to refine sequence patterns. 
Together, these advances not only establish a generalizable framework for efficient binder design against structurally large and otherwise inaccessible targets, but also support the broader “local-first” hypothesis as a guiding principle for PFNN-based design. 

\end{abstract}

\section{Introduction}

As a fundamental challenge in molecular engineering, \textit{de novo} design of protein binders not only holds significant therapeutic potential but also encapsulates understanding for the principles underlying protein folding and interaction \cite{gainza-cirauquiComputationalProteinDesign2018}. 
Recent advances in Protein Folding Neural Networks (PFNNs), most notably AlphaFold2 (AF2) \cite{jumperHighlyAccurateProtein2021}, have enabled considerable progress in computational binder design by streamlining the exploration of sequence and structural space and by improving the accuracy of \textit{in silico} validation \cite{watsonNovoDesignProtein2023,bennettImprovingNovoProtein2023}. 

However, as observed in large language models, output quality tends to degrade with increasing input size \cite{levySameTaskMore2024}; a similar behavior has been reported for protein folding networks \cite{changAlphaFold2KnowsProtein2024}. 
In binder design, this challenge has previously been sidestepped by trimming the input target to the smallest self-stabilizing and independently folding units, i.e., protein domains \cite{watsonNovoDesignProtein2023, pacesaBindCraftOneshotDesign2024, lvNovoDesignMiniprotein2024, ragotteDesignedMiniproteinsPotently2025}. 
Nevertheless, for binding surfaces located on large domains or spanning multiple domains, this strategy is inapplicable. 
In such cases, the low confidence of \textit{in silico} validation forces relaxation of selection thresholds \cite{watsonNovoDesignProtein2023}, while computational cost increases quadratically with input size, further limiting throughput and reducing the overall chance of success. 

To tackle these limitations, we inspect them through the lens of the energy function learned by PFNNs. 
While PFNNs are generally thought to capture an effective biophysical energy function of conformational space, with global constraints such as MSAs and templates serving mainly to accelerate conformational search rather than define the folding landscape \cite{roneyStateoftheartEstimationProtein2022}, it remains unclear to what extent this learned function diverges from the ground-truth global energy landscape. 
In fact, current PFNNs are trained with structures randomly cropped into equal-length fragments, irrespective of their global foldability, raising concerns about whether PFNNs truly capture global structural fitness.
Consistent with this concern, multiple lines of evidence indicate a strong local bias: (\textit{i}) in \textit{ab initio} folding, AF2 rapidly captures local residue–residue contacts, whereas remote contacts emerge much less frequently and only after extended iterations \cite{changAlphaFold2KnowsProtein2024}; (\textit{ii}) PFNN-hallucinated proteins tend to display limited conformational diversity, overly dense cores, elevated melting temperatures, reflecting over-optimization of local packing \cite{goverdeNovoProteinDesign2023}; and (\textit{iii}) in extreme cases, AF3 produces overlapping backbones with severe steric clashes to accommodate alternative binders, so long as local contacts remain satisfied \cite{jimenez-osesStructurePredictionAlternate2025}. 
Together, these findings suggest that the learned energy function is strongly biased toward local interactions, casting doubt on the indispensability of remote residues and global foldability in binder design. 

Motivated by this “local-first” hypothesis, we developed an “epitope-only” strategy, trimming the target domain to discontinuous surface residues surrounding the binding site and evaluating its performance in binder hallucination \cite{anishchenkoNovoProteinDesign2021, goverdeNovoProteinDesign2023}. 
The resulting “epitope-only” strategy significantly accelerated sampling and improved design quality, while preserving correlation with refold validation on the intact domain. 
Remarkably, this strategy enabled successful design for two targets previously considered intractable, ClpP and ALS3. 

Building on this strategy, we developed an efficient, extensible, and automated pipeline for mini-protein (MP) and cyclic-peptide (CP) binder design against large and rigid targets. 
The pipeline incorporates Monte Carlo (MC)–based evolutionary refinement to overcome local minima of gradient-based optimization, and ProteinMPNN redesign with position-specific bias to optimize local sequence features such as hydrogen-bond satisfaction \cite{fershtHydrogenBondingBiological1985} and isoelectric point (pI) balance \cite{guptaAntibodiesWeaklyBasic2022}. 
Compared with the original BindCraft pipeline, our modified framework achieved markedly higher sampling speed and success rate. 

Together, our findings not only reinforce the “local-first” mechanism suggested by AF2 but also provide a practical epitope-only strategy that substantially improves binder design efficiency. 
Moreover, our implementation offers an integrated platform for tackling large and challenging targets, thereby broadening the therapeutic scope of \textit{de novo} protein binders. 

\begin{figure}[htbp]
\centering  
\includegraphics[width=0.8\textwidth, keepaspectratio]{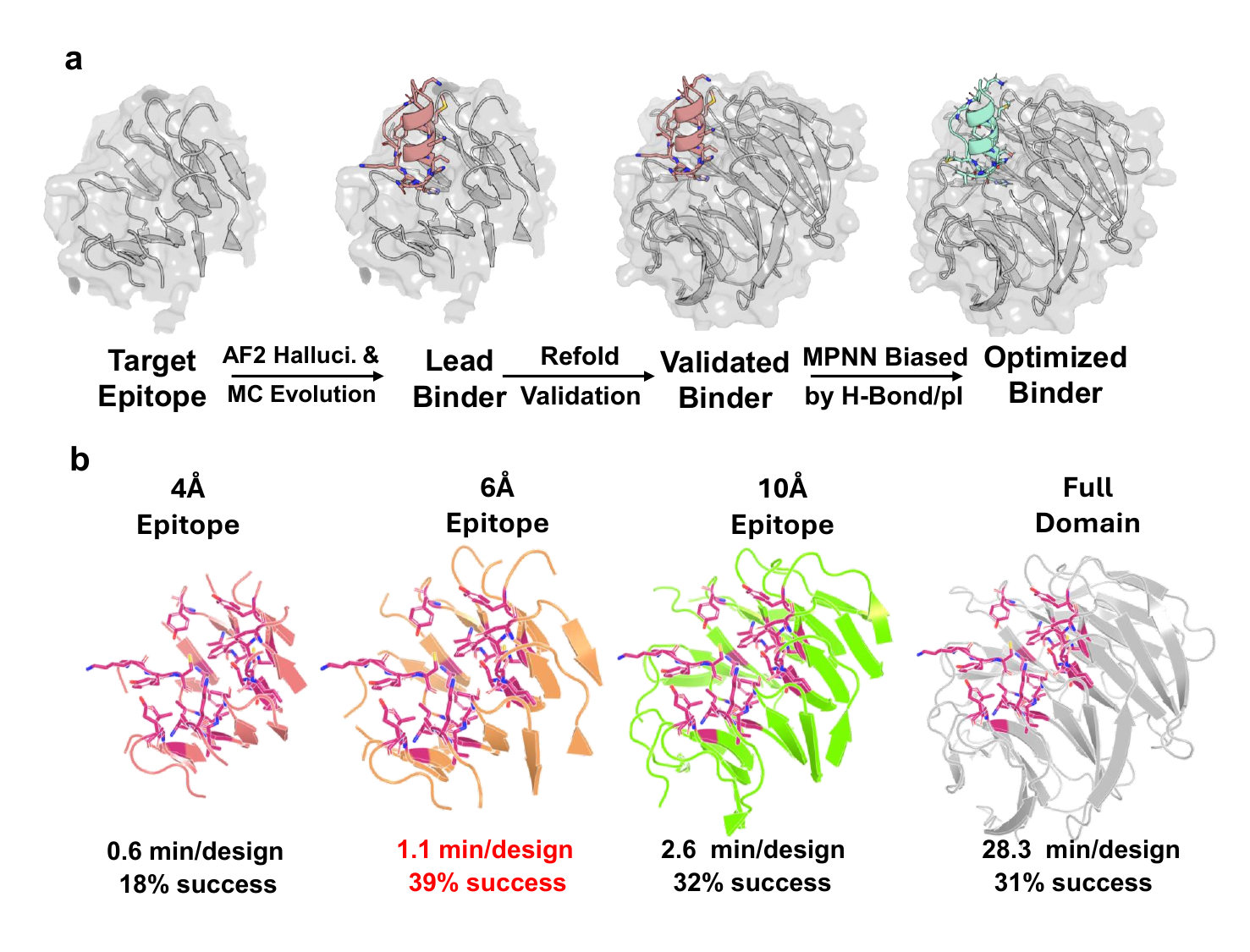}
\caption{
\textbf{Epitope-only hallucination for binder design against large targets. }
\textbf{a.} Workflow overview. Targets are cropped into discontinuous epitope for binder hallucination. Monte Carlo-based evolution is then performed to overcome local minima. 
Initial designs with high confidence are co-folded with intact target as validation. Eligible designs are redesigned with per-residue biased MPNN to improve developability and local features.
\textbf{b.} Increased speed/success rate of epitope-only hallucination strategy, showcased by CP binder design against WDR5. 
}
\label{fig:0-abstract}   
\end{figure}

\section{Related Work}
\subsection{Protein Folding Neural Networks (PFNNs)} 
The recent revolution in protein structure prediction was triggered by AlphaFold2 (AF2), a neural network that achieved unprecedented accuracy \cite{jumperHighlyAccurateProtein2021}. 
By integrating multiple sequence alignment (MSA) and structural template features with a novel invariant point attention module, AF2 and its analog RoseTTAFold \cite{baekAccuratePredictionProtein2021} enabled near-experimental accuracy in monomer folding for the first time. 
ESMFold further demonstrated that comparable accuracy can be achieved without querying MSAs, by leveraging a large-scale protein language model, while offering much faster inference \cite{linEvolutionaryscalePredictionAtomiclevel2023}. 
While multimer folding was initially pursued through finetuning AF2 on complex structures \cite{evansProteinComplexPrediction2022}, it was soon demonstrated that the original AF2 parameters trained only on monomers are already sufficient for multimer prediction \cite{gaoAF2ComplexPredictsDirect2022}. 
More recently, all-atom PFNNs such as AF3 \cite{abramsonAccurateStructurePrediction2024}, RF3 \cite{corleyAcceleratingBiomolecularModeling2025}, Chai-1 \cite{ChaidiscoveryChailab2024}, and Boltz-2 \cite{passaroBoltz2AccurateEfficient2025} have extended modeling to DNA, RNA, ligands, and ions, thereby enabling design and evaluation across a broader chemical space. 

\subsection{PFNN-derived Binder Evaluation} 
Beyond structure prediction, PFNNs also output a series of confidence scores, including point-wise (pLDDT), pair-wise (pAE), and global (pTM) measures. 
These scores have demonstrated state-of-the-art discriminative power for assessing structural quality and have been widely applied to evaluate designed proteins \cite{roneyStateoftheartEstimationProtein2022}. 
Nevertheless, their ability to distinguish true protein–protein interactions (PPIs) from artifacts remains limited: a retrospective study reported accuracies below 10\% \cite{bennettImprovingNovoProtein2023}. 
Moreover, this limitation persists even in the latest all-atom PFNNs, despite their substantially longer inference times \cite{manshourComprehensiveEvaluationAlphaFoldMultimer2024}. 
For these reasons, we focused our evaluations on AF2-like models, where inference is both efficient and well-validated. 

\subsection{PFNN-derived Binder Design} 
Several approaches have sought to repurpose PFNNs as generative models. 
One prominent direction is fine-tuning as diffusion models, exemplified by RFDiffusion \cite{watsonNovoDesignProtein2023}. 
Diffusion-based methods have proven effective for backbone generation, but co-design of sequence and structure remains challenging in this framework. 
In contrast, hallucination strategies allow simultaneous optimization of sequence and structure without additional training. 
By iteratively updating random sequences through gradient-based optimization of PFNN-derived confidence scores and geometric constraints, hallucination can directly produce tailored binder candidates \cite{anishchenkoNovoProteinDesign2021,goverdeNovoProteinDesign2023}. 

BindCraft (BC) represents the most systematic application of hallucination to binder design and has achieved remarkable \textit{in vitro} success \cite{pacesaBindCraftOneshotDesign2024}. 
BC employs AF2-multimer for hallucination \cite{evansProteinComplexPrediction2022} and AF2-monomer for refolding validation \cite{jumperHighlyAccurateProtein2021}, supplemented by several optimization and filtering steps. 
First, to alleviate the local minima that arise from gradient-based optimization, BC introduces a semi-greedy mutation procedure. 
Second, to address developability issues that often plague hallucinated binders \cite{raybouldFiveComputationalDevelopability2019,goverdeNovoProteinDesign2023}, BC applies ProteinMPNN-based redesign to refill non-interface residues. 
Finally, a suite of physics-based filters is used to eliminate designs with undesirable features, such as an excessive number of unsatisfied hydrogen-bond donors. 
While highly successful, these steps also reveal limitations: semi-greedy mutation can stagnate in suboptimal solutions, MPNN redesign is applied only in a global manner without fine-grained control of local sequence features, and efficiency remains limited when tackling large or multi-domain targets. 
These limitations motivate the methodological refinements explored in this work. 

\section{Rationale and Method}

\subsection{PFNN Energy Function Biased to Local Interaction }
Multiple sequence alignment (MSA) is widely recognized as a key determinant of folding accuracy \cite{elofssonProgressProteinStructure2023}, primarily by encoding co-evolutionary patterns that correlate with long-range contacts \cite{morcosDirectcouplingAnalysisResidue2011}. 
Although a deep MSA with accurate alignments and diverse sequences is generally regarded as beneficial for PFNNs, a growing body of work manipulating MSA has revealed counterintuitive behaviors. 
For example, end-to-end learned MSAs of relatively low quality can in fact improve folding accuracy \cite{pettiEndtoendLearningMultiple2023}, and sub-sampling MSAs has been shown to capture alternative conformations \cite{wayment-steelePredictingMultipleConformations2024, nunez-francoAlphaFold2PredictsAlternative2024}. 

In addition to MSA, global contacts may also be introduced through structural templates. 
In binder design, for instance, using the target structure as a template can enforce high-fidelity folding without requiring the computationally intensive construction of an MSA. 
Subsequent binder-target re-folding with the designed conformation as an ``initial guess'' can further preserve the binding pose while maintaining the discriminative power of confidence scores \cite{bennettImprovingNovoProtein2023}. 

Together, these observations underscore the heavy reliance of PFNNs on external global constraints. 
Nevertheless, PFNNs are still able to accurately rank candidate structures even in the absence of MSAs or templates \cite{roneyStateoftheartEstimationProtein2022}. 
It was thus deduced that PFNNs indeed learn an effective biophysical energy function, and that global constraints primarily serve to narrow down the search space rather than to define the folding landscape itself. 

However, given the residue cropping procedure in PFNN training—where sequences are randomly segmented into equal-length peptide fragments regardless of their overall foldability—it remains debatable whether the learned energy function truly captures global structural fitness. 
Indeed, recent evidence indicates that the learned function is strongly biased toward local interactions. 
For example, \textit{ab initio} folding with AF2 demonstrates that local residue–residue contacts are captured early and with high sensitivity, whereas remote contacts emerge less frequently and only after extensive iterations, implying that AF2 prioritizes local fitness \cite{changAlphaFold2KnowsProtein2024}. 
Similarly, hallucinated proteins—which reflect intrinsic PFNN preferences—tend to form densely packed cores and exhibit elevated melting temperatures (at the expense of solubility), pointing to an over-optimization of local interactions \cite{goverdeNovoProteinDesign2023}. 
This local bias is further highlighted by an extreme case in AF3, where the network produced overlapping backbones with severe steric clashes in order to accommodate two alternative binders at the same site—both of which can form ideal local contacts \cite{jimenez-osesStructurePredictionAlternate2025}. 

The concept of a ``local first'' mechanism was originally introduced to describe the \textit{ab initio} folding behavior of AF2 \cite{changAlphaFold2KnowsProtein2024}. 
Here, we extend this principle to the learned energy functions of PFNNs more broadly, arguing that these networks focus primarily on optimizing local interactions while displaying limited understanding of global structural fitness. 
This assumption forms the cornerstone of our epitope-only strategy. 

\subsection{``Epitope-only'' Strategy of Binder Design}
The local bias of the PFNN energy function underscores the importance of short-range interactions in both AF2 inference and hallucination. 
Furthermore, the training data of AF2-multimer are cropped by spatial (rather than sequential) vicinity, and thus deliberately disrupting structural completeness (\cite{evansProteinComplexPrediction2022}, Algorithm. 2), which further motivates our strategy to hallucinate binders against discontinuous epitope fragments. 

Our hallucination protocol is initialized following the same procedure as BC \cite{pacesaBindCraftOneshotDesign2024}. 
The input features of AF2 are repurposed for \textit{ab initio} folding of target–binder complexes. 
For the target, MSA-related features are initialized as zero matrices, while the target structure is parsed into template-related features. 
The binder is initialized as a random \verb+cluster_profile+ of size $[1, N_{b}, 23]$, where $N_{b}$ denotes the binder length and the last dimension corresponds to the amino acid type distribution at each residue. 
Binder sequences are generated from the \verb+cluster_profile+ via \verb+ArgMax+, and \verb+cluster_msa+ is constructed by repeating this sequence. 
At each step, the input features are passed through one randomly selected AF2-multimer model (out of five), and a composite loss function calculated by weighted-sum of the output confidence scores and geometric features (e.g. radius of gyration). 
Gradients are then backpropagated to the binder \verb+cluster_profile+ for optimization. 

The \verb+residue_index+ feature is leveraged to accommodate fragmented target structures. 
In AF2-like models, the \verb+residue_index+ $f^{r}$ is converted into a relative positional encoding $\mathbf{p}$ of size $[N_{\text{res}}, N_{\text{res}}]$, where $N_{\text{res}}$ denotes the total number of amino acids in the system and $\mathbf{p}_{ij} = f^{r}_{i} - f^{r}_{j}$ (clipped at $\pm 32$). 
By assigning residue indices from the fragmented target structure to $f^{r}$, AF2 can correctly process the relative positions between residues belonging to different segments. 
To represent binding surfaces spanning multiple chains, the fragments are treated as belonging to the same chain but separated by a gap of 50 in $f^{r}$. 

\subsection{MC-based Evolution}
Gradient-based hallucination is efficient but prone to being trapped in local minima. 
To escape such minima and explore a broader sequence space, the original BC employed semi-greedy (SM) evolution, in which random mutations are introduced and only accepted if they reduce the hallucination loss. 
While this strategy can improve local exploration, it remains susceptible to stagnation in suboptimal regions.

\begin{figure}[htp]
\centering  
\includegraphics[width=\textwidth, keepaspectratio]{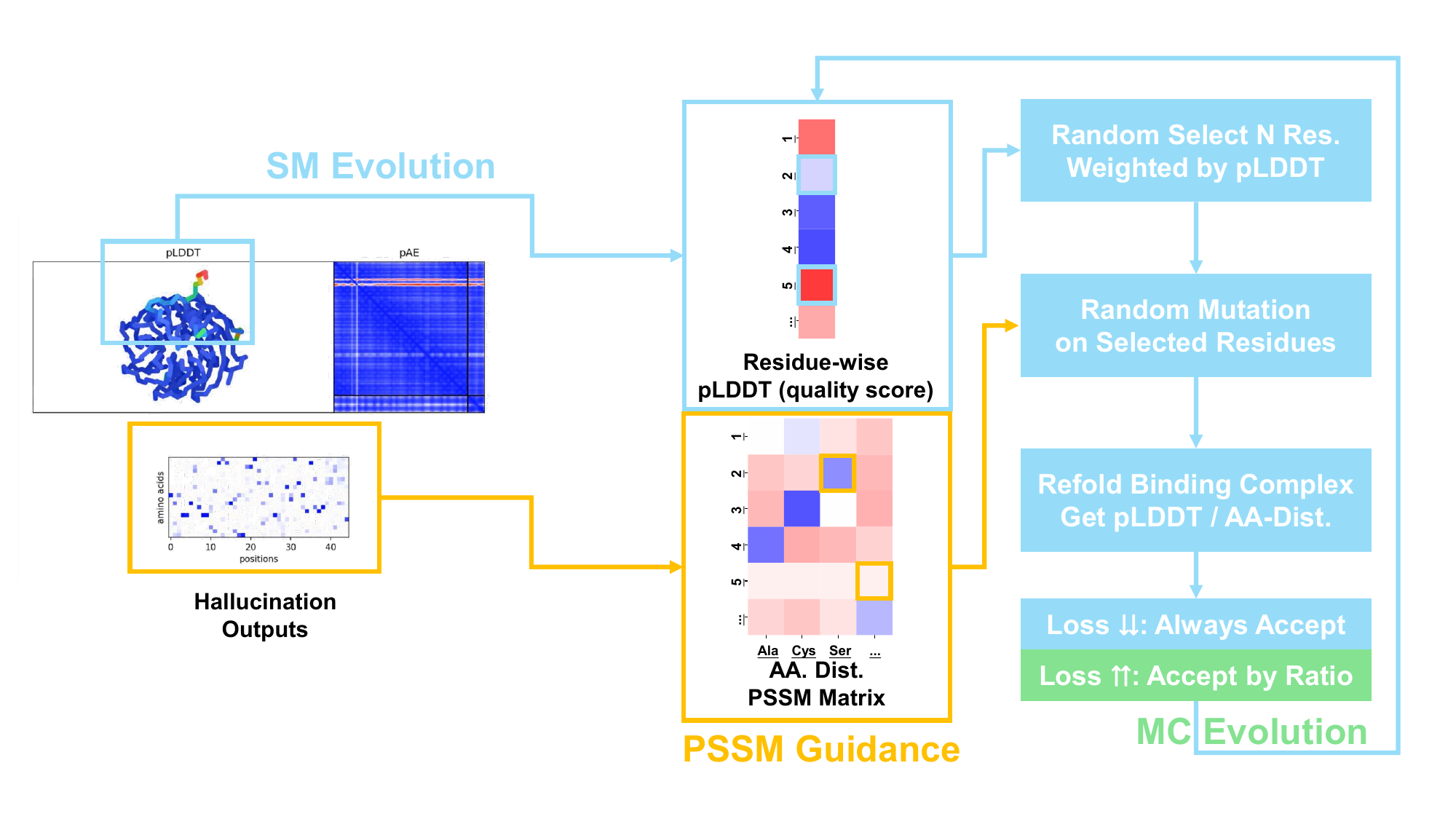}
\caption{
\textbf{Schema of Monte-Carlo evolution.}
At each step, we select 5\% of binder positions with probability $\propto$ 1-pLDDT and propose substitutions either uniformly or from a position-specific scoring matrix (PSSM) derived from the hallucinated amino acid distributions. After each proposal, we refold the complex and recompute pLDDT and the amino-acid distributions to update the hallucination loss. Unlike the semi-greedy scheme, MC also accepts loss-increasing moves, enabling early exploration then exploitation; we retain the lowest-loss frame along the trajectory. Matrix colors encode quality or sampling probability; red tones indicate lower quality/probability and blue tones higher.
}
\label{fig:s0-mc_schema}   
\end{figure}

To address this limitation, we replace SM with Markov chain Monte Carlo (MC)–based evolution (Fig. \ref{fig:s0-mc_schema}). 
Like SM, MC is performed immediately after hallucination and is driven by random residue mutations. 
At each step, 5\% of residues are randomly selected, with selection probability weighted by pLDDT such that low-confidence residues are more likely to be mutated. 
Selected residues are then substituted either uniformly among 20 amino acids (Random) or according to a position-specific scoring matrix normalized from the hallucinated amino acid distribution (PSSM). PSSM introduces bias guided by hallucination, potentially improving efficiency compared to purely random mutations.

The distinction between SM and MC lies in the acceptance criterion. 
SM accepts mutations only when the loss decreases, whereas MC may also accept unfavorable mutations with a probability determined by a simulated annealing scheme: 

\begin{equation}
P_{\text{accept}}(i, \Delta) = \exp\left(-\frac{\Delta}{T(i)}\right) 
= \exp\left(-\frac{\Delta}{T_{\text{init}} \cdot 2^{-i / \tau}} \right) 
= \exp\left(-\Delta \cdot \frac{2^{i/\tau}}{T_{\text{init}}} \right),
\end{equation}

\noindent where $\Delta = \text{loss}_{\text{new}} - \text{loss}_{\text{current}}$, 
$i$ is the current MC step, $T(i)$ is the exponentially decaying temperature, $T_{\text{init}}$ is the initial temperature, and $\tau$ is the half-life constant controlling the decay rate. 
As $i$ increases, $T(i)$ decreases, thereby lowering the probability of accepting loss-increasing mutations. 
A larger $\tau$ leads to slower cooling and thus a longer exploratory phase. 

For each trajectory, the best frame along the evolution is selected as the final design. 
For negative controls without any evolutionary refinement, the best hallucinated frame is taken as the final design.

\subsection{Biased MPNN Redesign}
Hallucinated binders often suffer from poor developability \cite{goverdeNovoProteinDesign2023}. 
To improve sequence quality, the original BC pipeline employs ProteinMPNN inverse folding \cite{dauparasRobustDeepLearning2022} to resample binder residues outside the protein–protein interface (PPI). 
ProteinMPNN infers position-specific amino acid distributions from the $C_\alpha$ coordinates of the complex and the amino acid identities of the target, PPI residues, and previously sampled binder residues, and then autoregressively fills the masked binder sequence. 
BC additionally introduced simple global bias to exclude cysteine, which otherwise complicates experimental handling. 

However, global biases cannot address local sequence features that are critical for binding and developability. 
For example, unoccupied polar residues on the PPI—those not forming hydrogen bonds or salt bridges with the target—can reduce binding affinity by $\sim$3 kcal/mol (up to $\sim$160-fold) \cite{fershtHydrogenBondingBiological1985}. 
Likewise, excessive charged residues on the non-PPI surface can skew the isoelectric point (pI), leading to aggregation or non-specific interactions \cite{guptaAntibodiesWeaklyBasic2022}. 
In BC, designs with such unfavorable local patterns were simply discarded, which is wasteful given the scarcity of qualified designs for difficult targets. 

\begin{figure}[htp]
\centering  
\includegraphics[width=\textwidth, keepaspectratio]{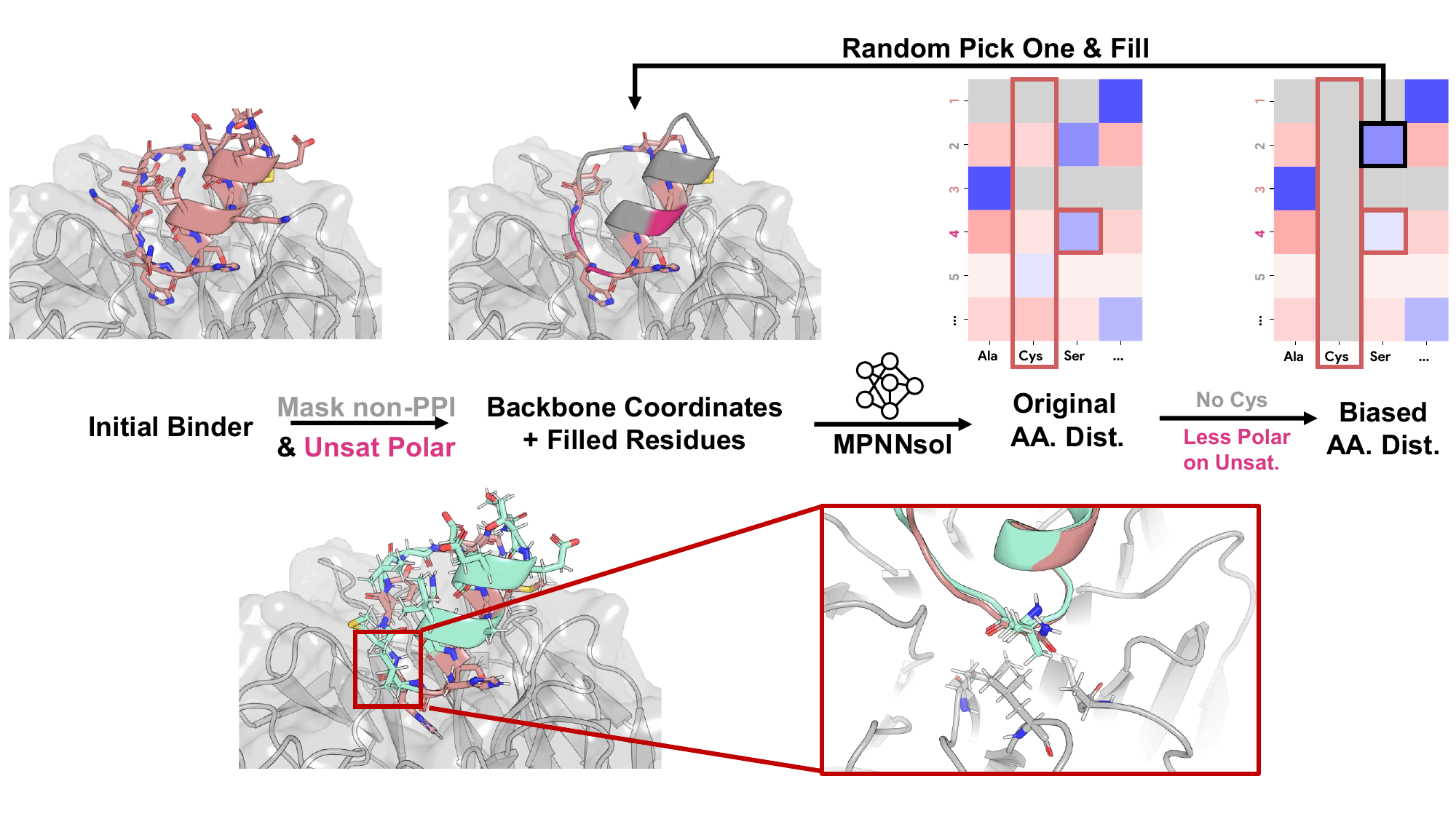}
\caption{
\textbf{Schema of MPNN redesign with per-residue bias.}
\textbf{(Top)} For each hallucinated binder, we mask non-interface residues and optionally polar interface sites that lack an intermolecular H-bond or salt bridge after relaxation (grey). We apply a global exclusion for cysteine (logit = $-10^6$) and mild penalties ($-ln4$) to disfavor polar/charged residues at (i) unsatisfied polar PPI positions (hydrophilic-occupancy bias) and (ii) solvent-exposed non-PPI positions with rSASA > 0.5 (surface-charge bias, omitted for simplicity). Matrix colors encode ProteinMPNN sampling probability; red tones indicate lower probability and blue tones higher. 
\textbf{(Bottom)} Redesigned protein (teal), generated with per-residue-biased ProteinMPNN, eliminates unsatisfied hydrogen bonds with the target.
}
\label{fig:s0-mpnn_schema}   
\end{figure}

Here we introduce MPNN redesign with per-residue bias for better control of local patterns (Fig. \ref{fig:s0-mpnn_schema}), with which we explicitly optimized these local features: 
1) Hydrophilic occupancy bias: residues on the PPI that are polar but unpaired in hydrogen bonds or salt bridges are included in redesign, with polar/charged residues penalized. 
2) Surface charge bias: charged residues on non-PPI solvent-exposed surfaces (relative SASA $>$ 0.5) are penalized to maintain a pI value favorable for therapeutic developability \cite{guptaAntibodiesWeaklyBasic2022}. 
Bias is introduced by adding a logit penalty of $-10^{6}$ to exclude cysteine (baseline), and $-\ln 4$ for specific residues disfavoring as in \cite{goverdeComputationalDesignSoluble2024}. 




\section{Experimental Study}

\subsection{Set-up}
\begin{figure}[h]
\centering  
\includegraphics[width=0.9\textwidth, keepaspectratio]{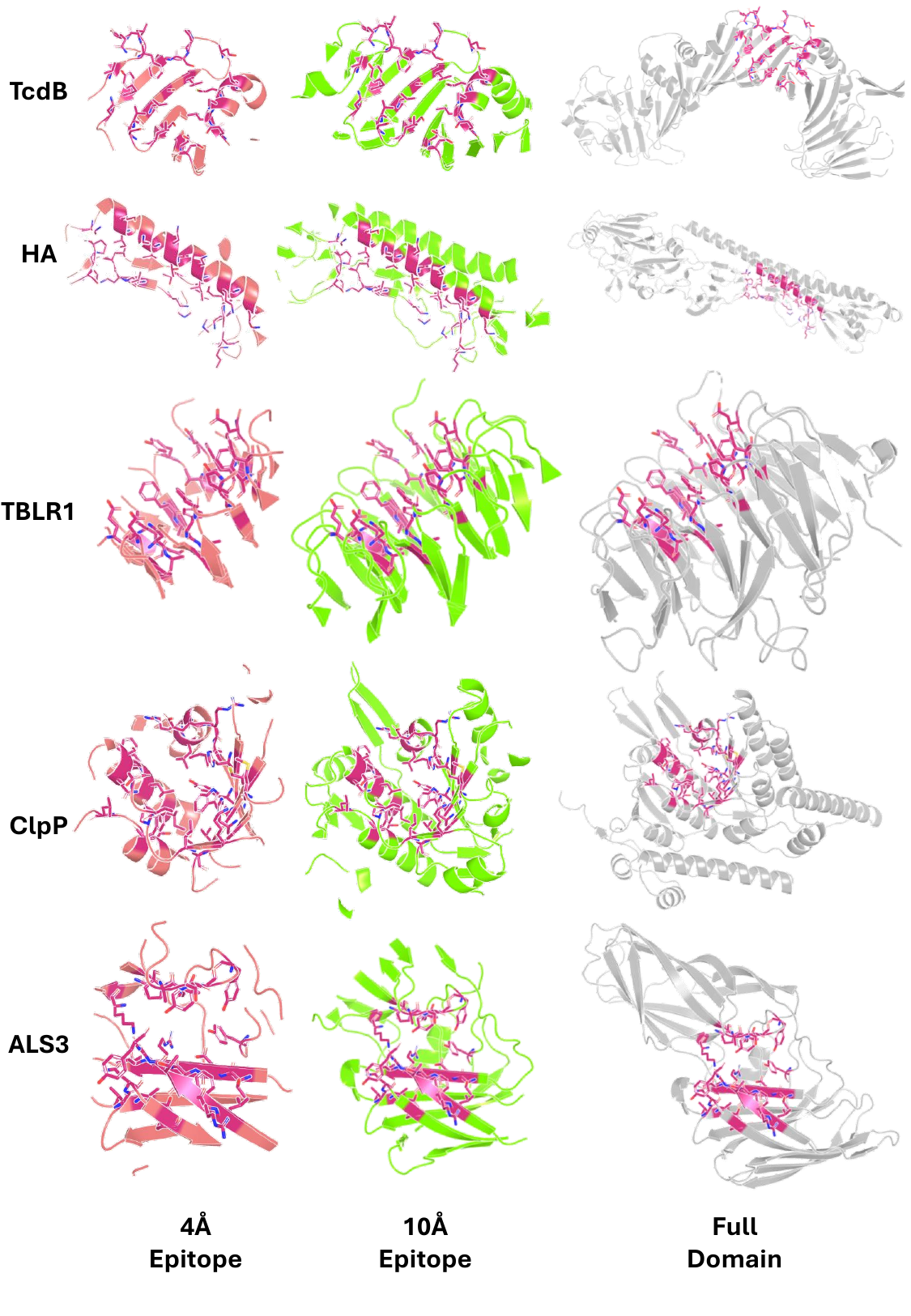}
\caption{
\textbf{Overview of tested targets, hot spots and epitopes}. Hot spots are highlighted in warm pink and side chain conformation. Epitope regions are selected by distance to hot spots and shown in the left/right (with different distance cutoff). Full domain, as the minimum foldable unit where epitopes exists, are shown in the right.}
\label{fig:s0-epitope_view}   
\end{figure}

We evaluated our approach (Fig. \ref{fig:0-abstract}a) on six biological systems with large or multi-domain binding surfaces (Tab. \ref{Table:target_metadata}). 
To avoid arbitrariness, epitopes were defined strictly by the distance to functional hot spots, i.e., residues in contact with known binders in crystal structures (Fig. \ref{fig:0-abstract}b\&\ref{fig:s0-epitope_view}). 
To ensure broad applicability, both mini-protein (MP) designs against flat surfaces and cyclic-peptide (CP) designs against narrow pockets (enabled by cyclic positional offsets \cite{rettieCyclicPeptideStructure2023}) were tested. 

300 binders are Hallucinated for each target. By default, 15 steps of MC evolution without PSSM guidance is performed. For fair comparison, design quality was first assessed after hallucination–evolution with AF2-multimer. 
Successful designs,defined as those with pLDDT > 0.8 and interface pAE (i-pAE) < 0.35, were then refolded with intact target domains using AF2-monomer. 
To facilitate folding of these large multimeric systems, the designed binding pose was provided as the “initial guess” \cite{bennettImprovingNovoProtein2023} and the initial frame coordinates (Suppl. Alg. 20 in \cite{jumperHighlyAccurateProtein2021}). 
Binders that achieved both high refold confidence (pLDDT>0.8 and i-pAE<0.35) and consistency with the design (binder RMSD < 3.5\r{A}) were considered successful refolds, i.e., successful binders. 

For MPNN redesign, the top 100 initial designs (lowest i-pAE) were selected for further optimization. 
From each design, 1600 sequences were sampled, clustered into five groups by sequence similarity (spectral clustering, scikit-learn \cite{scikit-learn}), and the highest-scoring sequence in each group was selected for refold validation. 
Hydrophilic-occupancy bias is applied to both MP and CP. Inter-molecular H-bonds and salt-bridges are identified with geometric constraint by MdAnalysis \cite{gowers2019mdanalysis} after relaxation \cite{conwayRelaxationBackboneBond2014}
Surface-charge bias is only applied to MP. pI is calculated by PROPKA3 \cite{olssonPROPKA3ConsistentTreatment2011}. 
Additional success criteria were applied according to the specific bias: $7.5 < \text{pI} < 9$ for surface-charge bias, and fewer than three unsatisfied donors together with more than three satisfied donors for hydrophilic-occupancy bias, following previous study \cite{pacesaBindCraftOneshotDesign2024, guptaAntibodiesWeaklyBasic2022}. 

Finally, overall efficiency of the proposed pipeline was compared against the original BC pipeline, using the full set of BC filters as the evaluation criterion. 

\subsection{Epitope-Only Hallucination Improves Speed and Success} 
\begin{figure}[htp]
\centering  
\includegraphics[width=\textwidth, keepaspectratio]{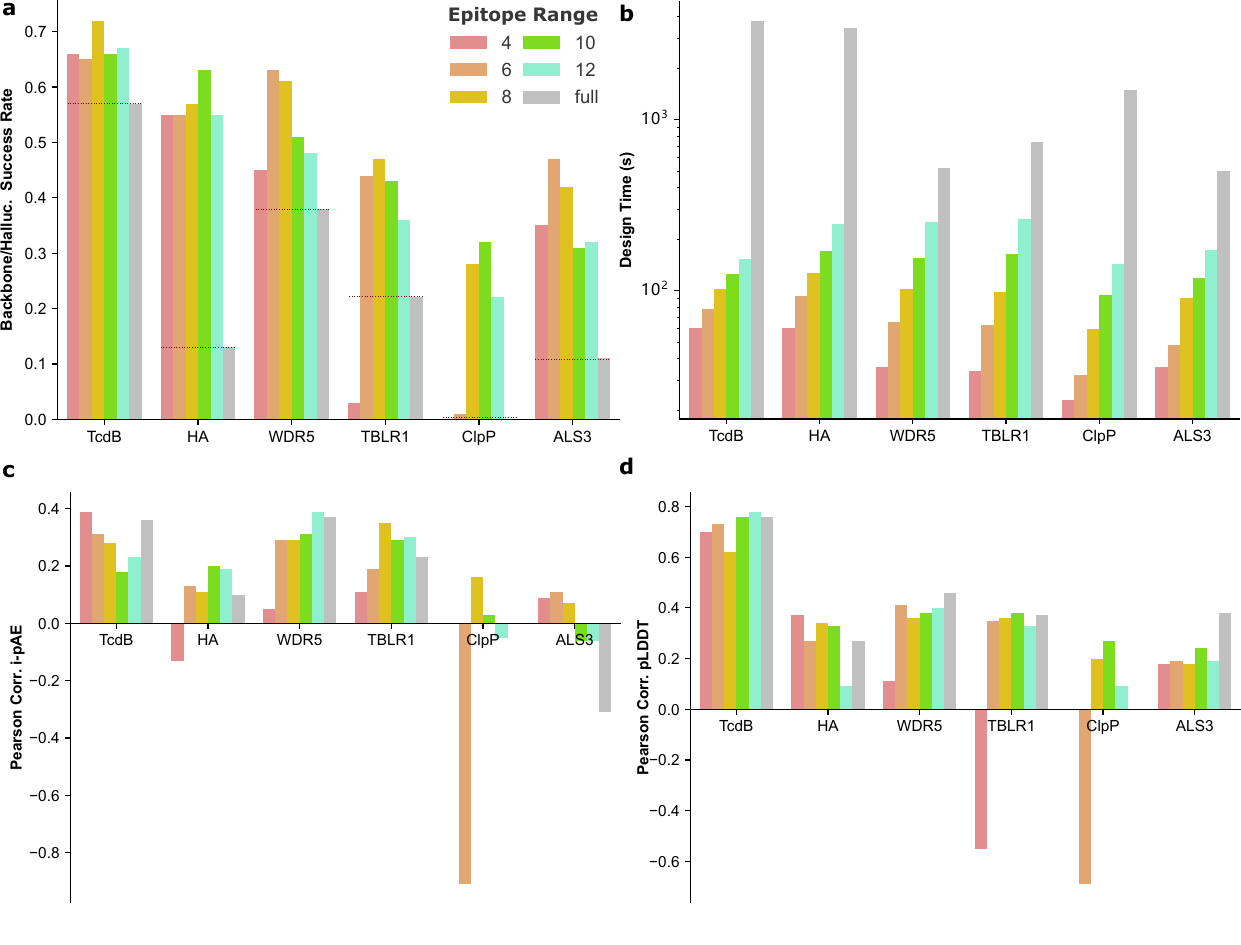}
\caption{
\textbf{Ext. efficiency improvement with epitope-only hallucination Strategy. }
\textbf{a.} increased success rate in initial design / refold validation compared to full domain as input. Red dashed line indicates the performance of full domain. 
\textbf{b.} reduced per-design time. per-refold-success missing bar indicates no success design, thus no available data. 
\textbf{c-d.} correlation of confidence scores between design and refold. See Fig. \ref{fig:s1-plddt-corr}-\ref{fig:s1-ipae-corr} for details. 
}
\label{fig:s1-epi_results}   
\end{figure}

As expected, sampling speed increased markedly as the epitope range was narrowed (Fig. \ref{fig:s1-epi_results}a). 
Design success rates also improved steadily (Fig. \ref{fig:s1-epi_results}b), consistent with previous reports that AF2 performance declines with increasing system size \cite{changAlphaFold2KnowsProtein2024}.  

\begin{figure}[tp]
\centering  
\includegraphics[width=\textwidth, keepaspectratio]{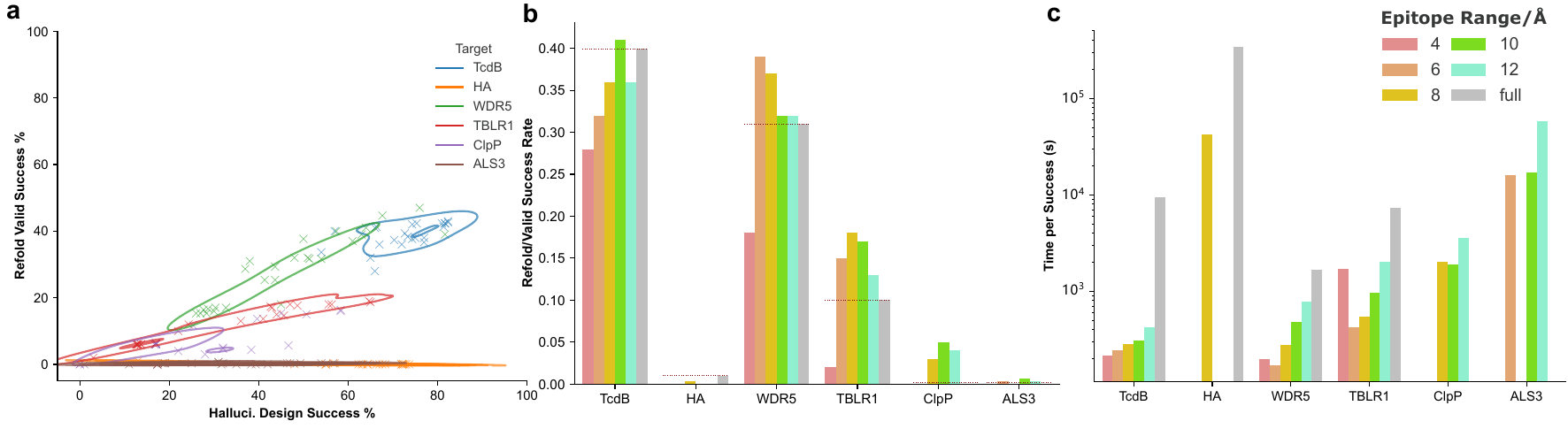}
\caption{
\textbf{Efficiency improvement with epitope-only hallucination strategy. }
\textbf{a.} Correlation between Design/Refold Success. Each dot represents a batch of design with different conditions (e.g., target epitope range, evolution conditions). Lines are kernel density estimate (KDE) plot of dots distribution.
\textbf{b.} Increased success rate in refold validation compared to full domain as input. Red dashed line indicates the performance of full domain. 
\textbf{c.} Reduced per-refold-success time. Missing bar indicates no success design, thus no available data. 
}
\label{fig:1-epi}   
\end{figure}

\begin{figure}[htp]
\centering  
\includegraphics[width=\textwidth, keepaspectratio]{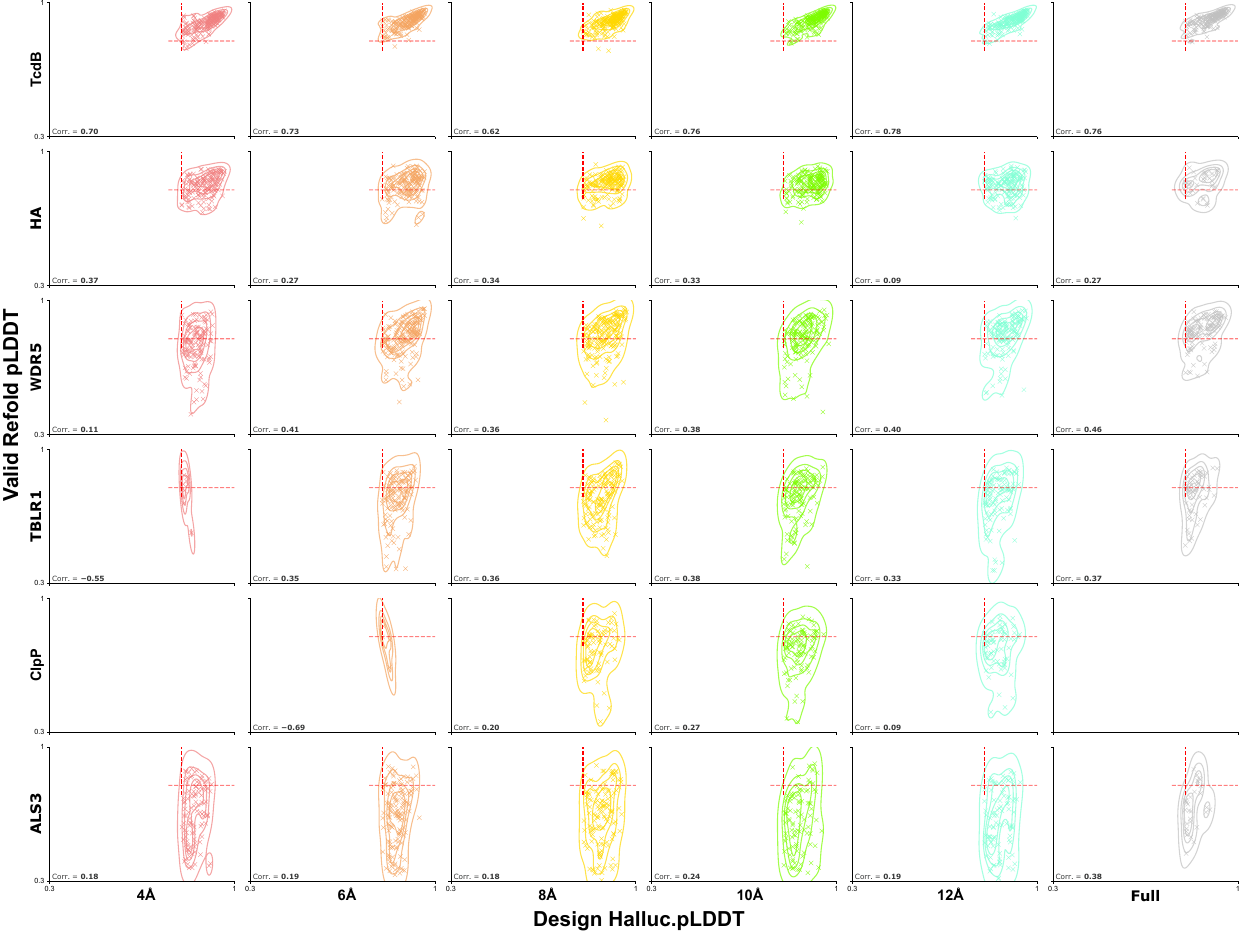}
\caption{
\textbf{Correlation between design and refold pLDDT}, as supplement to Fig. \ref{fig:s1-epi_results}c. Dashed line indicates the acceptable thresholds. Hallucination is conducted with AF2-multimer model \#1-5 on epitopes or full domain. Refold validation is conducted with AF2-monomer model \#1-2 on full domain.  
}
\label{fig:s1-plddt-corr}   
\end{figure}

\begin{figure}[htp]
\centering  
\includegraphics[width=\textwidth, keepaspectratio]{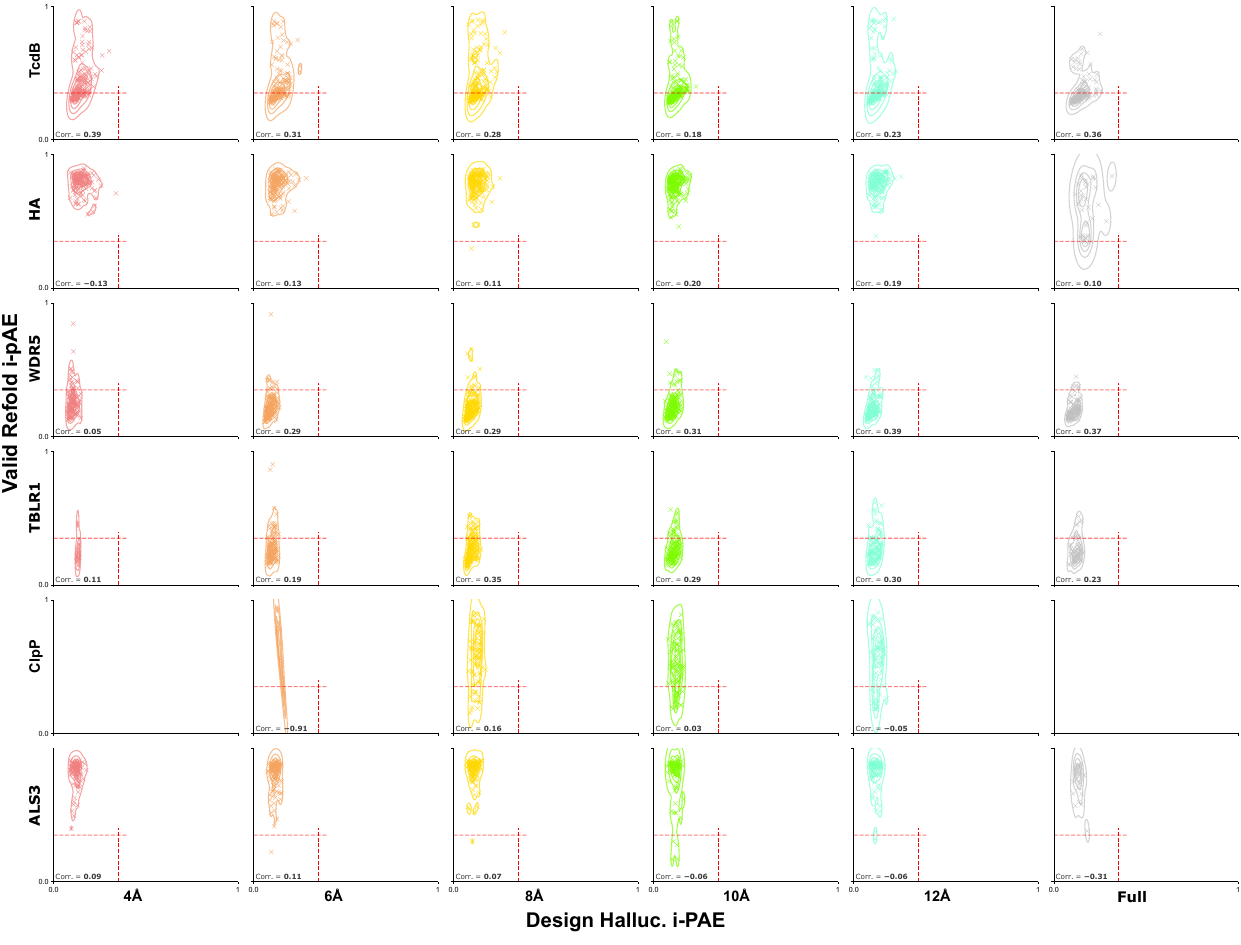}
\caption{
\textbf{Correlation between design and refold i-pAE}, as supplement to Fig. \ref{fig:s1-epi_results}d. Dashed line indicates the acceptable thresholds. Hallucination is conducted with AF2-multimer model \#1-5 on epitopes or full domain. Refold validation is conducted with AF2-monomer model \#1-2 on full domain. 
}
\label{fig:s1-ipae-corr}   
\end{figure}

Importantly, reducing the epitope size did not compromise the correlation between design and refold confidence scores (Fig. \ref{fig:1-epi}a, \ref{fig:s1-plddt-corr}–\ref{fig:s1-ipae-corr}), indicating that distal regions of the target protein are largely redundant and can even be confounding for binder design. 
As a consequence, the epitope-only strategy also improved refold success (Fig. \ref{fig:1-epi}b). 
For relatively easy targets (baseline full-domain success >30\%), the gains were modest (TcdB +3\%, WDR5 +19\%). 
For more difficult cases, however, the improvements were dramatic: +80\% for TBLR1, and a shift from essentially “undesignable” to substantially designable for ALS3 and ClpP. 
The only exception was HA, where success did not improve; this outcome can be attributed to extremely poor consistency between design and refold (Fig. \ref{fig:1-epi}a), suggesting that factors other than domain size underlie its difficulty.  

By combining faster sampling with higher-quality designs, overall sampling efficiency—defined as the average time per successful design—was enhanced across all six targets, with improvements ranging from 3.5-fold to 44.2-fold. 
These results highlight the robust and consistent benefits of the epitope-only strategy. 

\subsection{MC Evolution Rescues Poor Hallucination} 
\begin{figure}[tp]
\centering  
\includegraphics[width=0.8\textwidth, keepaspectratio]{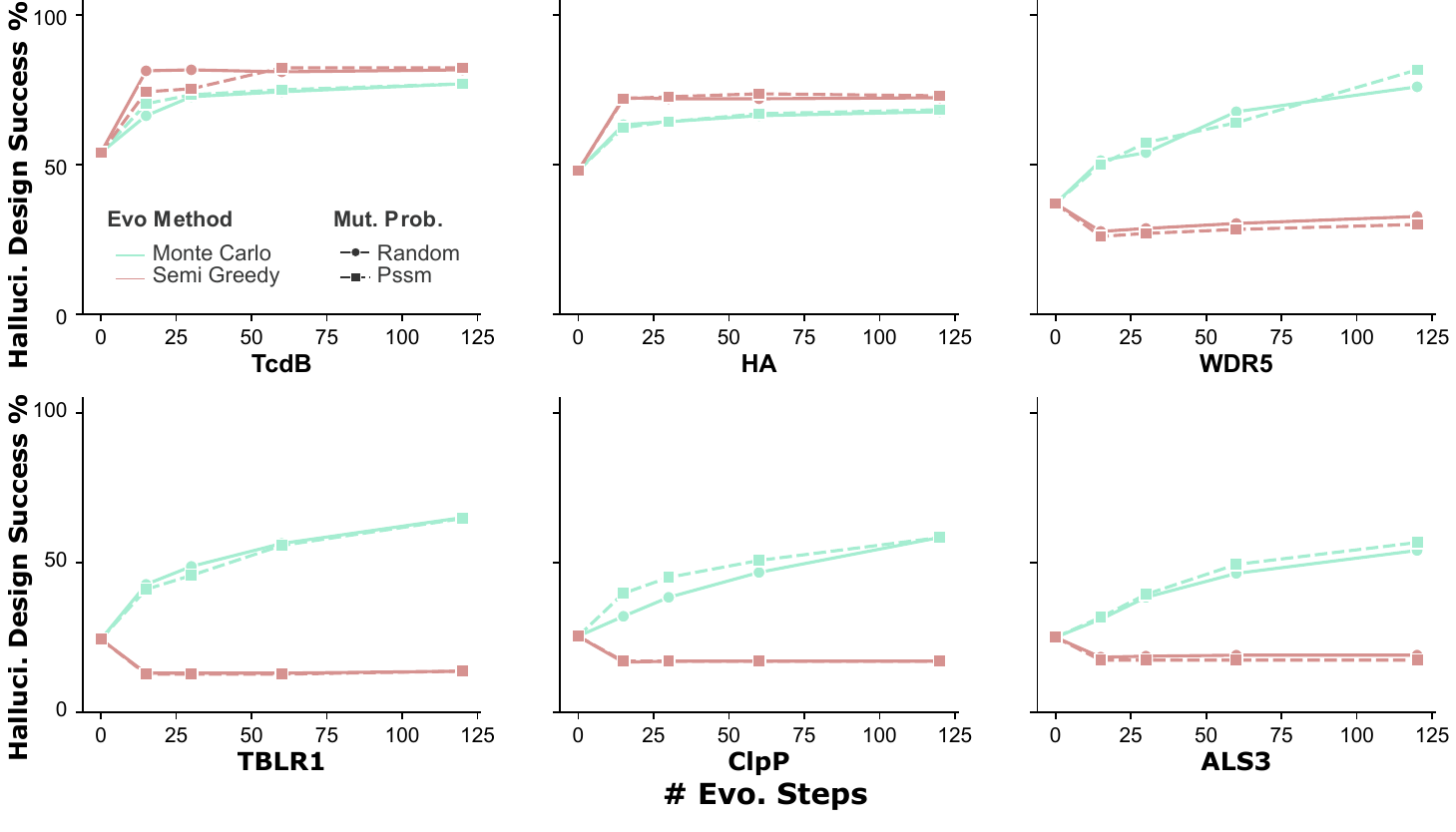}
\caption{
\textbf{Design success rate improved with MC evolution.}
}
\label{fig:2-mc-design}   
\end{figure}

Our results show that semi-greedy (SM) evolution is only effective when the initial hallucination result is already of reasonable quality (Fig. \ref{fig:2-mc-design}, targets TcdB and HA). 
Even in these cases, backbone quality quickly plateaus after ~30 steps. 
Incorporating PSSM guidance as the mutation probability provides only marginal benefit, likely because it restricts exploration to the hallucinated sequence distribution—contrary to the intended purpose of evolution. 
As a result, SM, with or without PSSM, performs poorly for difficult targets.  

In contrast, for systems where SM offers little advantage, MC evolution markedly improved success rates: +130\% for WDR5, +364\% for TBLR1, +241\% for ClpP, and +184\% for ALS3 (relative to SM). 
For easier systems where SM suffices, MC caused only modest reductions (-7\% for TcdB, -6\% for HA). 
Notably, MC performance continued to improve with extended sampling and showed no saturation even at 120 steps, suggesting substantial untapped potential. 

\begin{figure}[tp]
\centering  
\includegraphics[width=\textwidth, keepaspectratio]{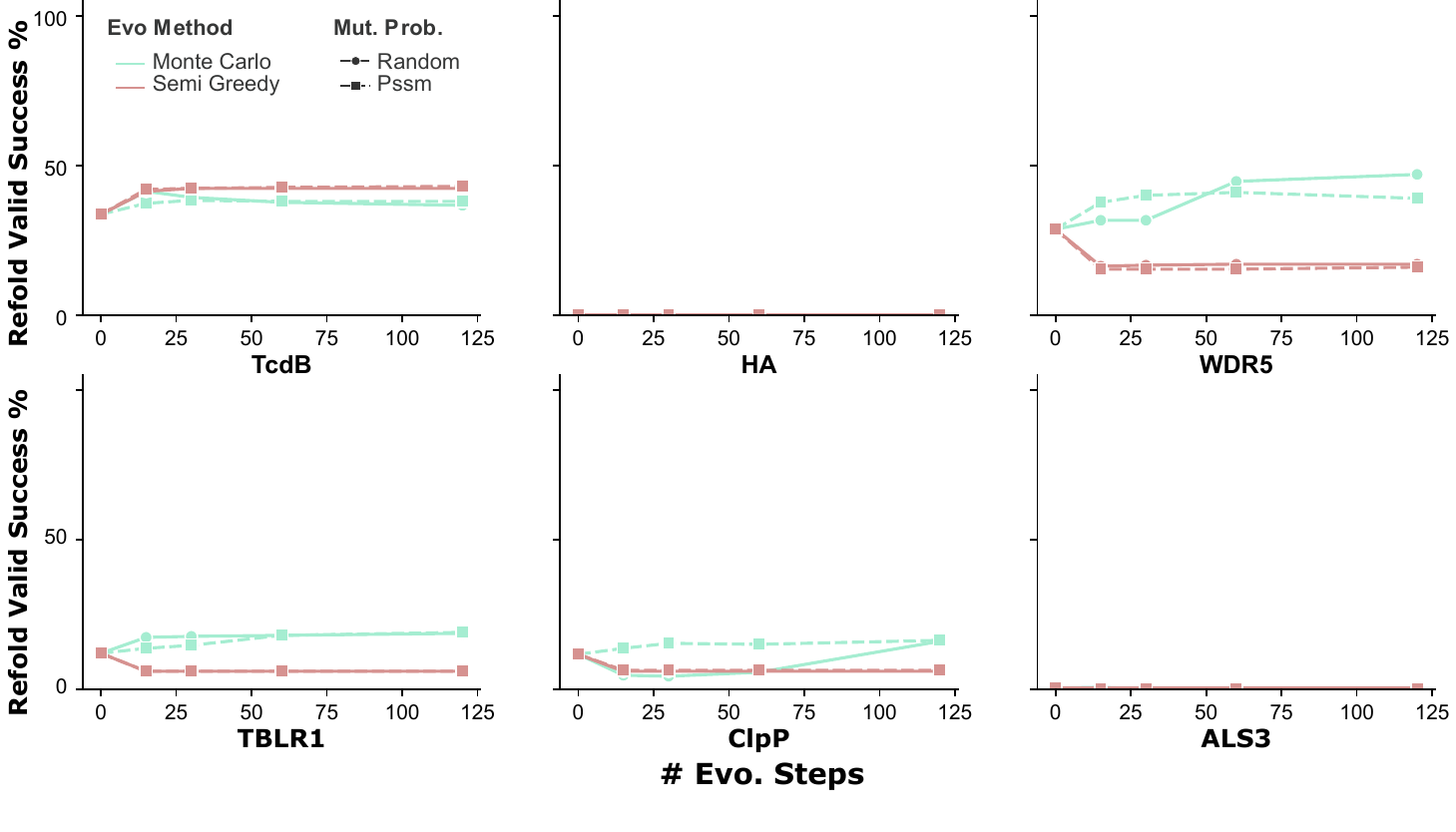}
\caption{
\textbf{Refold success rate improved with MC evolution.} 
}
\label{fig:s2-evo_steps-refold}   
\end{figure}

Admittedly, an improved design success rate does not necessarily translate into higher \textit{in silico} yields (Fig. \ref{fig:s2-evo_steps-refold}) despite a strong correlation observed (Fig. \ref{fig:1-epi}a). 
Nevertheless, because the hallucinated conformation also serves as the refolding template, enhancing its quality is essential for reliable refold evaluation.

\subsection{Biased MPNN Enhances Local Sequence Features} 

\begin{figure}[tp]
\centering  
\includegraphics[width=\textwidth, keepaspectratio]{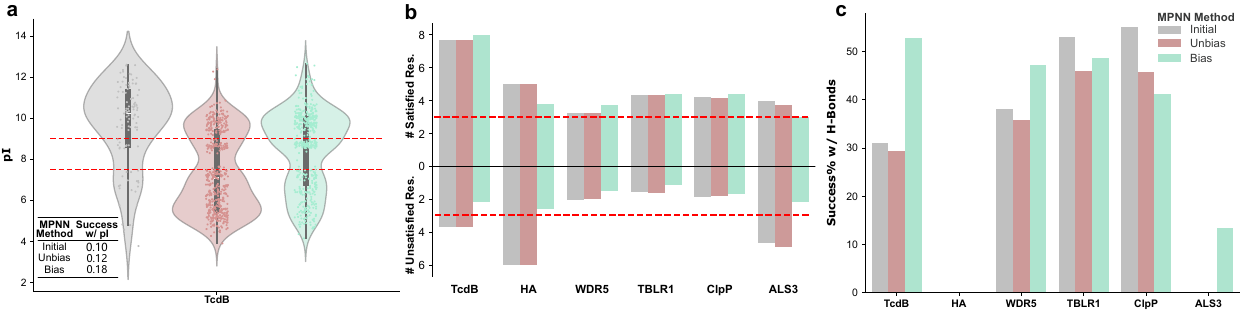}
\caption{
\textbf{Optimized sequential features with biased MPNN}.
\textbf{a.} Optimized distribution of pI in TcdB MP binders. Dashed line indicates the acceptable thresholds. Inner table reported the refold success rate with extra pI constraint. 
\textbf{b.} Optimized Occupation of polar residue on PPI.  Dashed line indicates the acceptable thresholds. 
\textbf{c.} Refold success rate with extra H-bond criteria. 
}
\label{fig:3-mpnn}   
\end{figure}

For the surface-charge bias task, the proportion of eligible designs increased from 10\% to 20\% with biased MPNN, whereas the default MPNN achieved only 13\% (Fig. \ref{fig:3-mpnn}a). 
When combined with refolding quality, this translated into an overall success rate improvement from 12\% to 18\%.  

For the hydrophilic-occupancy task, biased MPNN markedly reduced the number of unsatisfied polar residues in targets where this was a severe issue (Fig. \ref{fig:3-mpnn}b, TcdB, HA, ALS3). 
For WDR5 and TBLR1, where the average changes were modest, the fraction of binders meeting the occupancy criterion still rose substantially from 38\% to 51\%, suggesting real gains that were masked by already-sufficient designs (Fig. \ref{fig:3-mpnn}c). 
admittedly, as a trade-off, introducing penalties slightly reduced refold confidence, and in cases where occupancy was not a bottleneck (e.g., ClpP), the overall success rate decreased from 46\% to 41\%. 

Taken together, these results show that biased MPNN provides a generally effective means of improving local sequence features across both tasks, particularly when the targeted feature represents a limiting factor for binder quality. 
However, its benefits are less pronounced—or even counterproductive—when the targeted feature is not a bottleneck.  

\subsection{Enhanced Performance with Full Pipeline} 
Due to low efficiency of original BC pipeline, only full-pipeline baseline on TcdB and TBLR1 is done. 
As shown in Tab. \ref{Tab: Overall_Success}, our modified pipeline has achieved >4 times acceleration and 3 to 4 times higher success rate.

\section{Discussion}
Motivated by the hypothesis of a “local-first” mechanism in PFNN energy functions, our epitope-only hallucination strategy achieved substantial improvements in both the speed and quality of binder design. 
Building on this foundation, our optimized pipeline—which further incorporates MC-based evolution and biased MPNN redesign to address common pitfalls in designing against large, difficult targets—appears promising for unlocking the therapeutic potential of previously intractable systems.

At the same time, our findings should be interpreted in light of several limitations. 
Larger-scale retrospective analyses are required to confirm the fidelity of folding confidence scores between epitope-only and full-domain settings. 
Additional support for the “local-first” hypothesis could also be obtained through gradient analyses of distal target residues relative to binder profiles. 
Since our current results are based on AF2-multimer, whose spatial-cropping training scheme may itself contribute to the observed performance, it will also be important to test the strategy on AF3-like all-atom PFNNs. 

Finally, several limitations of our experimental setup should be acknowledged. 
First, additional design tasks are required for comprehensive evaluation, with particular emphasis on mini-protein benchmarks, given their relatively weaker performance of MC-based evolution. 
Second, the modest reduction in refold confidence observed with biased MPNN redesign likely results from suboptimal penalty settings rather than an inherent limitation of the approach. 
Moreover, experimental validation will ultimately be essential to assess the extent to which the observed computational gains translate into functional binders. 
Future work will therefore focus on broadening the scope of design tasks, refining methodological parameters, and incorporating wet-lab validation to rigorously evaluate the robustness of the proposed pipeline.

\newpage
\bibliographystyle{nips} 
\bibliography{references}  

\newpage
\section*{Appendix}

\newpage

\begin{tabularx}{\textwidth}{Y Y Y Y Y Y Y} 
\caption[Metadata of protein target]{
\textbf{Metadata of protein target.} \textbf{MP}, mini-protein. \textbf{CP}, cyclic peptide. \textbf{*}full ClpP target is defined as its dimerization form, which is modeled by Boltz-2 \cite{passaroBoltz2AccurateEfficient2025} with 6BBA as template. 
}\label{Table:target_metadata}\\ 
\toprule
\textbf{Target} &\textbf{ PDB ID} & \textbf{Binder Modality} & \textbf{Domain Size} & \textbf{Target Chain} & \textbf{Binder Chain} \\ 
\midrule
\textbf{WDR5}   & 2G99   & CP   & 304           & B            & C    \\      
\textbf{TBLR1}  & 5NAF   & CP  & 348            & B            & F     \\     
\textbf{ClpP}   & 6BBA\textbf{*}  & CP  &  390         & A,B          & L      \\   
\textbf{ALS3}   & 4LEB   & CP  & 326         & A            & B    \\     
\textbf{TcdB}   & 6C0B   & MP   & 506       & A            & B   \\       
\textbf{HA}     & 5VLI   & MP    & 495      & A,B          & C   \\     
\bottomrule
\end{tabularx}


\newpage

\newpage

\newpage
\clearpage


\begin{tabularx}{0.6\textwidth}{Y Y Y c}
\caption{
\textbf{Speed and quality improvement of the whole Pipeline.}
BC, the original BindCraft pipeline. 
EC, our pipeline, tentatively termed EpitopeCraft. 
}\label{Tab: Overall_Success} \\
\toprule
\textbf{Target} & \textbf{Pipeline} & \textbf{Sampl. Time (s)} & \textbf{Success Rate} \\
\midrule
\multirow{2}{*}{\textbf{TcdB}} & BC & 1440 & 0.07 \\
 & EC & 337 & 0.29 \\
\midrule
\multirow{2}{*}{\textbf{TBLR1}} & BC & 1337 & 0.05 \\
  & EC & 300 & 0.14 \\
\bottomrule

\end{tabularx}

\end{document}